\documentclass[twocolumn,showpacs,preprintnumbers,amsmath,amssymb,aps,prb]{revtex4}

\usepackage{graphicx}
\usepackage{dcolumn}
\usepackage{bm}
\usepackage{subfigure}
\usepackage{natbib} 
\usepackage{multirow}

\begin{document}

\title{Photoinduced filling of near nodal gap in Bi$_2$Sr$_2$CaCu$_2$O$_{8+\delta}$}
 
\author{Z. Zhang$^{1}$, C. Piovera$^{2}$,  E. Papalazarou$^{3}$, M. Marsi$^{3}$, M. d'Astuto$^{1}$, C. J. van der Beek $^{1}$, A. Taleb-Ibrahimi${^4}$, and L. Perfetti$^{2}$}

\affiliation{$^{1}$ Institut de Min\'eralogie, de Physique des Mat\'eriaux,
et de Cosmochimie (IMPMC), Sorbonne Universit\'es - UPMC Univ Paris 06,
case 115, 4, place Jussieu, 75252 Paris cedex 05, France
}

\affiliation{
$^{2}$ Laboratoire des Solides Irradi\'{e}s, Ecole Polytechnique, CNRS, CEA, Universit\'e Paris-Saclay, 91128 Palaiseau, France}

\affiliation{$^{3}$ Laboratoire de Physique des Solides, CNRS, Univ. Paris-Sud, Universit\'e Paris-Saclay, 91405 Orsay Cedex, France}

\affiliation{
$^{4}$ Synchrotron SOLEIL, L'Orme des Merisiers, Saint-Aubin-BP 48, F-91192 Gif sur Yvette, France}

\date{\today}

\begin{abstract}

We report time and angle resolved spectroscopic measurements in optimally doped Bi$_2$Sr$_2$CaCu$_2$O$_{8+\delta}$. The photoelectron intensity maps are monitored as a function of temperature, photoexcitation density and delay time from the pump pulse. We evince that thermal fluctuations are effective only for temperatures near to the critical value whereas photoinduced fluctuations scale linearly at low pumping fluence. The minimal energy to fully disrupt the superconducting gap slightly increases when moving off the nodal direction. No evidence of a pseudogap arising from other phenomena than pairing has been detected in the explored region of reciprocal space. On the other hand, an intermediate coupling model of the photoinduced phase transition can constistently explain the gap filling in the near as well as in the off-nodal direction. Finally, we observed that nodal quasiparticles develop a faster dynamics when pumping the superconductor with fluence large enough to induce the total collapse of the gap.

\end{abstract}


\maketitle

\section{Introduction}

The superconductors of the copper-oxide family have been matter of extensive investigations and are still subject of fierce debates. After 30 years of research, some issues have been settled, whereas others remain controversial. The evolution of the superconducting order parameter with temperature and doping level is an exemplary case. It is nowadays well established that the single particle gap near to the nodal direction is an hallmark of superconductivity. By moving half a way from the nodal direction it possible to observe that a remnant pairing persists up to temperatures higher than the critical value $T_c$  \cite{Campuzano, Kaminski, Shin}. This fluctuating regime can be explained by intermediate coupling models \cite{Perali} with a finite rate of pair-breaking \cite{Norman, Dessau, Shin}. Finally, it has been established that the antinodal region displays an additional pseudogap that interplays and, eventually, competes with superconductivity \cite{Kaminski,Shen}.

In the last years, the field has been enriched by experimental protocols that are capable of detecting the single particle spectra out of equilibrium conditions \cite{Perfetti_ARPES, Smallwood_science, Current}. Smallwood \emph{et al.} reported the collapse and subsequent recovery of the single particle gap after photoexcitation by a short laser pulse \cite{Smallwood_science, Smallwood_gap}. Their data indicate that the gap is more robust when moving towards the antinodes. Moreover, the minimal fluence necessary to melt the near nodal gap has been related to a qualitative change in the dynamics of nodal Quasi-Particles (QPs). Soon after, Ishida \emph{et al.} reproduced the gap melting, observed Bogoliubov excitations and outlined a concurrent reduction of QPs coherence \cite{Ishida}. Apparently, the recovery of the gapped phase proceeds within several picoseconds, due to electronic cooling by phonon emission. On this same timescale, the dynamics of a reforming condensate also affects the electromagnetic response function \cite{Demsar,Kaindl} and the transient population of hot electrons \cite{Smallwood_science}. Indeed, the cooling of QPs displays a dramatic slow down when the system enters in the superconducting phase \cite{Piovera,Kirchmann}. Such effect persists at photoexcitation fluence much larger than the threshold value necessary for the complete disruption of superfluid density. As a consequence, we proposed that a remnant pairing may protect QPs from energy dissipation. Alternatively, Smallwood et al. developed a model that would explain the slow cooling of QPs as the result of a dynamical gap opening \cite{Smallwood_Model}.

This article reports additional measurements on the photoinduced collapse of the superconducting gap in optimally doped Bi$_2$Sr$_2$CaCu$_2$O$_{8+\delta}$ (Bi2212). We aim to a comparative study of the Cooper pairs melting by thermal excitations and ultrafast laser pulses. In both cases the near nodal gap is filled by the appearance of a fluctuating condensate. The threshold fluence necessary to completely fill the gap depends slightly on the azimuthal angle and becomes larger towards the antinodes. The section of Brillouin zone that can be explored by a photon source centered at 6.3 eV allow us to cover roughly half of the Fermi surface. By extending recent experiments\cite{Smallwood_gap} to higher pumping fluence, we exclude that a pseudogap survives to high photoexcitation density. Instead, the gradual filling of the gap can be accurately reproduced by an intermediate coupling model accounting for the finite rate of pair-breaking  \cite{Perali}.

The second novelty of our work is the comparison of the transient state following photoexcitation with an adiabatic heating across the phase transition. On one hand, we find that photoinduced and thermal fluctuations result in similar collapse of the gapped spectral function. On the other hand the evolution of the superconducting state clearly depends on the applied perturbation. Thermal fluctuations are effective only for temperatures near to the critical value whereas photoinduced fluctuations scale linearly with the pumping fluence. The peculiar behavior of the photoexcited state suggests that Cooper pairs scatter with a non-thermal distribution of excited phonons.

At last, the fluence dependence of the gap is compared with the energy dissipation of low energy QPs. We use 600 fs pulses to perform high resolution spectroscopy of the superconducting gap and 80 fs pulses to accurately follow the QP dynamics. The data show that nodal QPs develop a fast relaxation component at the same threshold fluence where the gap collapses \cite{Smallwood_QP}. These results are discussed in terms of a dynamical gap opening \cite{Smallwood_Model}.

The article is organized as follow: section \ref{sec2} reports the experimental configurations employed from time resolved ARPES measurements, sections \ref{sec3} and \ref{sec4} focus on the dependence of the near nodal gap on temperature and pumping fluence, section \ref{sec5} extends the equilibrium and non-equilibrium spectroscopy to the off nodal direction, section \ref{sec6} includes a general discussion of the experimental data and the comparative analysis of the gap evolution with the energy dissipation of nodal QPs.

\begin{figure}
\begin{center}
\includegraphics[width=1\columnwidth]{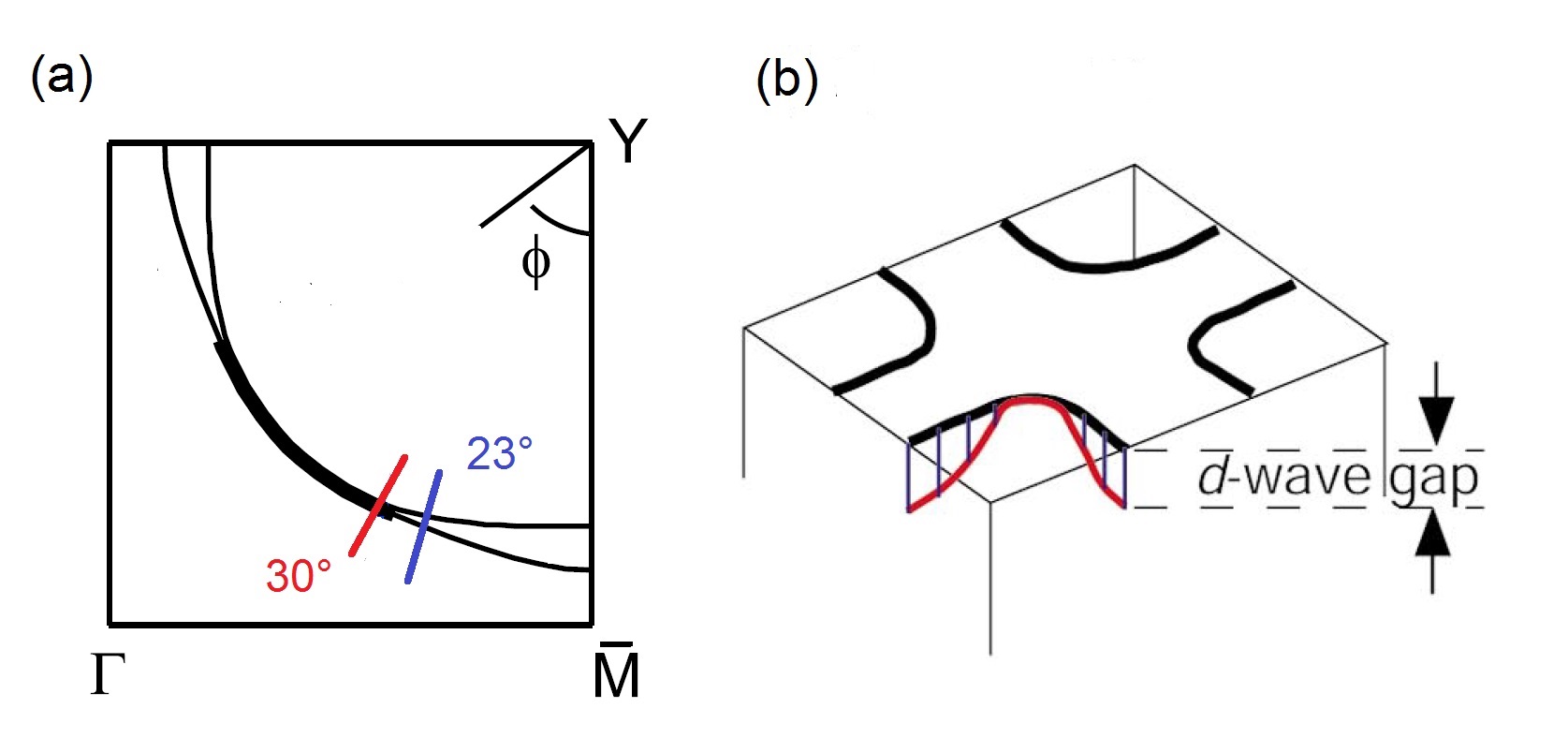}
\caption{(a) Fermi surface of Bi2212 in the first Brillouin zone. The red and blue line indicate cuts in reciprocal space where time resolved measurements have been performed. (b) Superconducting gap with $d$-wave symmetry}
\label{Fig1}
\end{center}
\end{figure}

\section{Methods}\label{sec2}

We investigate single crystals of optimally doped Bi$_2$Sr$_2$CaCu$_2$O$_{8+\delta}$ (Bi2212) with transition temperature $T_c=91$ K. Time resolved photoemission experiments are performed on the FemtoARPES setup \cite{FemtoARPES}, using a Ti:sapphire laser system delivering 35 fs pulses at 1.55 eV (780 nm) with 250 kHz repetition rate. Part of the laser beam is used to generate 6.3 eV photons through cascade frequency mixing in BBO crystals. The 1.55 eV and 6.3 eV beams are employed to photoexcite the sample and induce photoemission, respectively. Two different setups have been optimized for complementary experiments on the electrons dynamics. The quasiparticles relaxation has been probed with short pulses having duration of 80 fs and bandwidth of 70 meV. These pulses can follow the temporal evolution of excited electrons with high accuracy but do not provide the required energy resolution to monitor the superconducting gap. As a consequence, high resolution measurements have been done with UV pulses of 600 fs duration and spectral bandwidth below 15 meV. In each case the field amplitude of the probe has been reduced until space charge distortions dropped below the detectable limit. The incident photoexcitation fluence on the sample is evaluated by imaging the pump and probe beam in the focal plane. All reported measurements have been performed on freshly cleaved crystals at the base pressure of $7\times10^{-11}$ mbars.

\section{Near nodal gap in Equilibrium}\label{sec3}

\begin{figure}
\begin{center}
\includegraphics[width=1\columnwidth]{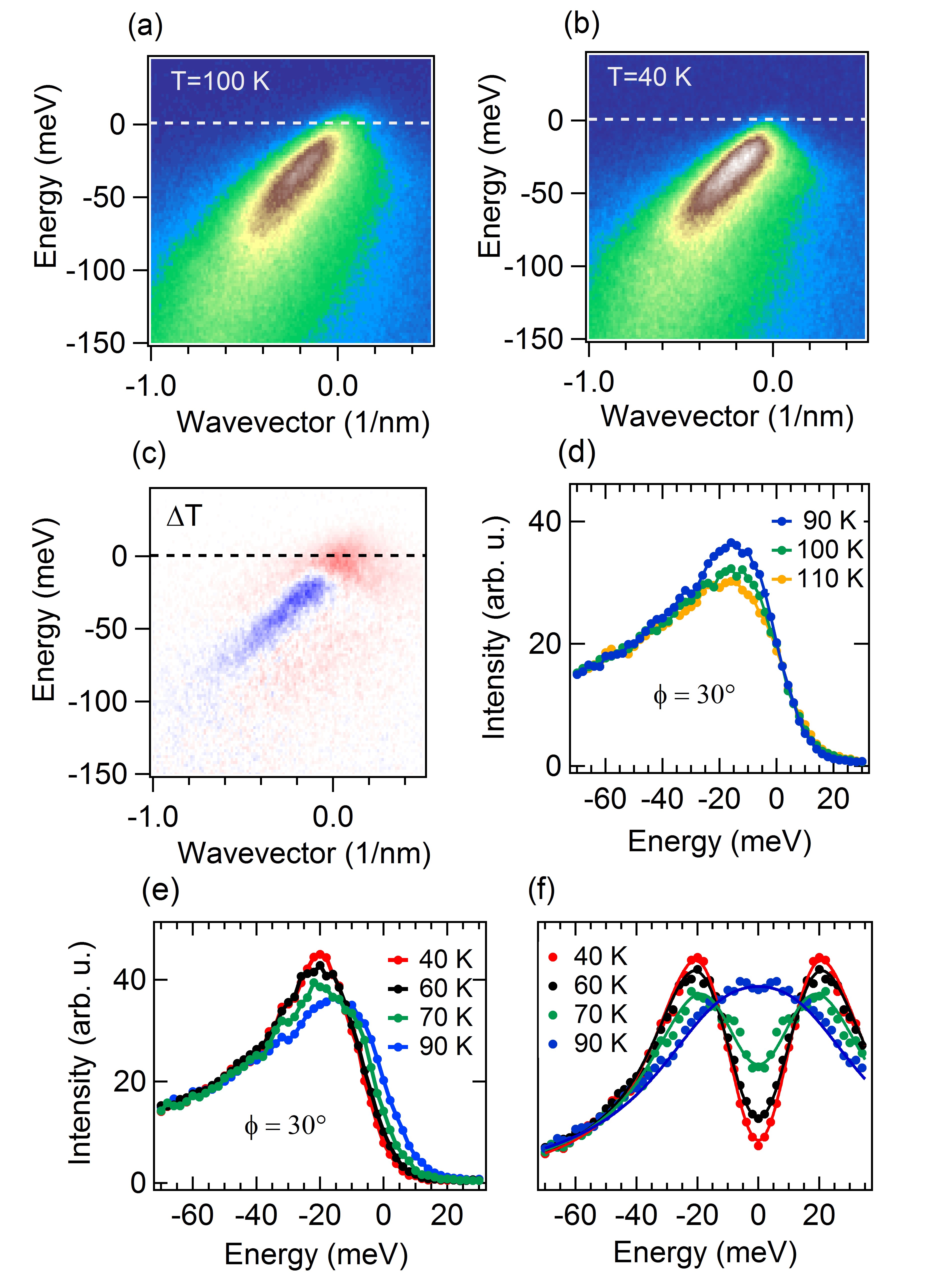}
\caption{The data in this image have been acquired in equilibrium conditions along a direction cutting the Fermi surface at azimuthal angle $\phi=30^\circ$. Photoelectron intensity maps at 100 K (a) and 40 K (b). (c) Differential signal between the intensity map collected at 100 K and 40 K. Energy distribution curves extracted at the Fermi wavevector for several temperatures above (d) and below $T_c$ (e). (f) Symmetrized EDCs extracted at the Fermi wavevector for different temperatures. Fitting curves (solid lines) based on the self energy  $\Sigma(k,\omega)=-i \gamma+\Delta^2/(\omega+\epsilon_k+i\gamma)$ are superimposed to experimental data (marks).}
\label{Fig2}
\end{center}
\end{figure}

\begin{figure} 
\begin{center} 
\includegraphics[width=1\columnwidth]{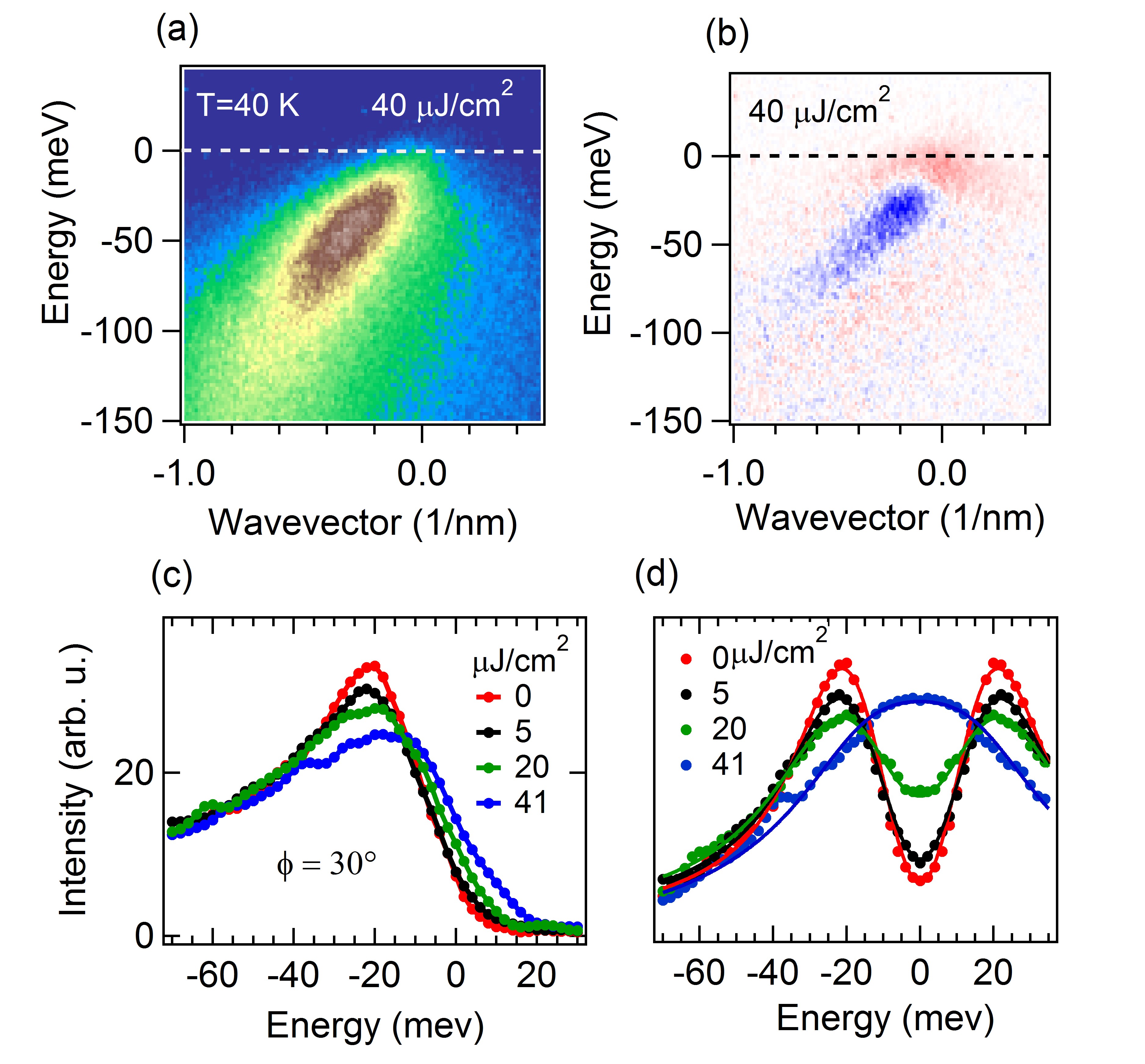} 
\caption{The data in this image have been acquired at azimuthal angle of $30^\circ$ and base temperature of 40 K for different values of photoexitation fluence. (a) Photoelectron intensity map collected 0.6 ps after the arrival of a  pump pulse depositing 40 $\mu J/cm^2$. (b) Pump-on minus pump-off intensity difference upon photoexictation with 40 $\mu J/cm^2$. EDCs (c) and symmetrized EDCs (d) extracted at the Fermi wavevector for different pumping fluence. Fitting curves (solid lines) are superimposed to symmetrized EDCs (marks).}
\label{Fig3}
\end{center} 
\end{figure}

As shown in Fig. \ref{Fig1}, we define the zero of the azimutal angle as the $\bar{M}-Y$ direction in reciprocal space, so that the nodal point is at $\phi=45^\circ$. First we analyze the development of a near nodal gap for azimuthal angle $\phi = 30^\circ$. Figure \ref{Fig2}(a) shows the ARPES intensity maps acquired at the equilibrium temperature of 100 K. The QP peak gains coherence and intensity for excitation energy above -70 meV. This energy scale is related to collective modes coupling to single-particle excitations and coincides to a kink in the QP dispersion \cite{Damascelli}. Figure \ref{Fig2}(b) shows the intensity map collected after cooling the sample in the superconducting phase at 40 K. In contrast to the previous case, a small but detectable gap inhibits the quasiparticle crossing of the Fermi level. We show in Fig. \ref{Fig2}(c) the difference intensity map between data acquired at 100 K and 40 K. Such plot visualize the redistribution of spectral weight upon thermal melting and serves as reference for the data on the photoinduced phase transition.

The gradual formation of the superconducting gap is obtained by acquiring intensity maps at several intermediate temperatures. Figure \ref{Fig2}(d,e) show the Energy Distribution Curves (EDCs) extracted at the Fermi wavevector for each one of these maps. The EDCs are normalized only by the total acquisition time, making possible a direct comparison of the relative intensity between different curves. As shown in figure \ref{Fig2}(d), the peak of the EDCs merely loose intensity as long as the temperature is above the critical value. Instead, the shift of leading edge observed in Fig. \ref{Fig2}(e) is the hallmark of an electronic gap developing below $T_c$. Following a common procedure in data treatment \cite{Shin,Kaminski,Smallwood_gap, Campuzano}, we show in Fig. \ref{Fig2}(f) the symmetrized EDCs at different equilibrium temperatures. The distance of the peaks from the Fermi level is a phenomenological indicator of gap magnitude and attains the value of $\cong20$ meV at 40 K.

As originally proposed by Norman et al. \cite{Norman}, we model the spectral function at the Fermi wavevector by the phenomenological self energy $\Sigma(k,\omega)=-i \gamma+\Delta^2/(\omega+\epsilon_k+i\gamma)$, where $\epsilon_k$ is the band dispersion, $\Delta$ is superconducting gap and $\gamma$ is the pair-breaking rate. 
Perali \emph{et al.} have shown that such phenomenological expression is a reasonable approximation as long as the asymmetry of the spectral function is not too large \cite{Perali}. In the intermediate coupling regime the pair-breaking rate $\gamma$ increases subtantially near the transition temperature and overcomes $\Delta$ in the normal phase. The resulting spectral function $A(k,\omega)=-\frac{1}{\pi}$Im$(1/(\omega-\epsilon_k-\Sigma))$ is convoluted in energy and momentum to account for the finite experimental resolution. Finally, a smooth background has been included to reproduce the incoherent spectral weight adding up at high energy. As shown by Fig. \ref{Fig2}(f), the curves obtained by such self energy fit accurately the experimental data. The fitting parameters of Fig. \ref{Fig4}(a) indicate that $\Delta$ displays minor variations for $T\leq T_c$. As expected, the superconducting phase transition is ruled by the sudden increase of the pair-breaking rate $\gamma$ when the system approaches the critical point. The gap is completely filled once $\gamma$ overcomes the value of $\Delta$. We stress that this scenario is different from the weak coupling BCS limit. In the latter case the temperature window where fluctuations become visible is not detectable and the pair-breaking rate is negligibly small.

\section{Near nodal gap in photoexcited samples}\label{sec4}

Next we investigate the evolution of the near nodal gap upon sudden photoexcitation with 1.5 eV pulses. Figure \ref{Fig3}(a) shows photoelectron intensity maps acquired at the azimuthal angle of 30 degrees, equilibrium temperature of 40 K and incident fluence of $ F= 40 \mu J/cm^2$. The images have been collected for a delay time corresponding to the maximal pump-probe signal. We show in \ref{Fig3}(b) difference intensity maps between the pump-on and pump-off case. Strong similarities between Fig. \ref{Fig3}(b) and Fig. \ref{Fig2}(c) indicates that photoexcitation and thermal fluctuations result in comparable transfer of spectral weight.

\begin{figure} 
\begin{center} 
\includegraphics[width=1\columnwidth]{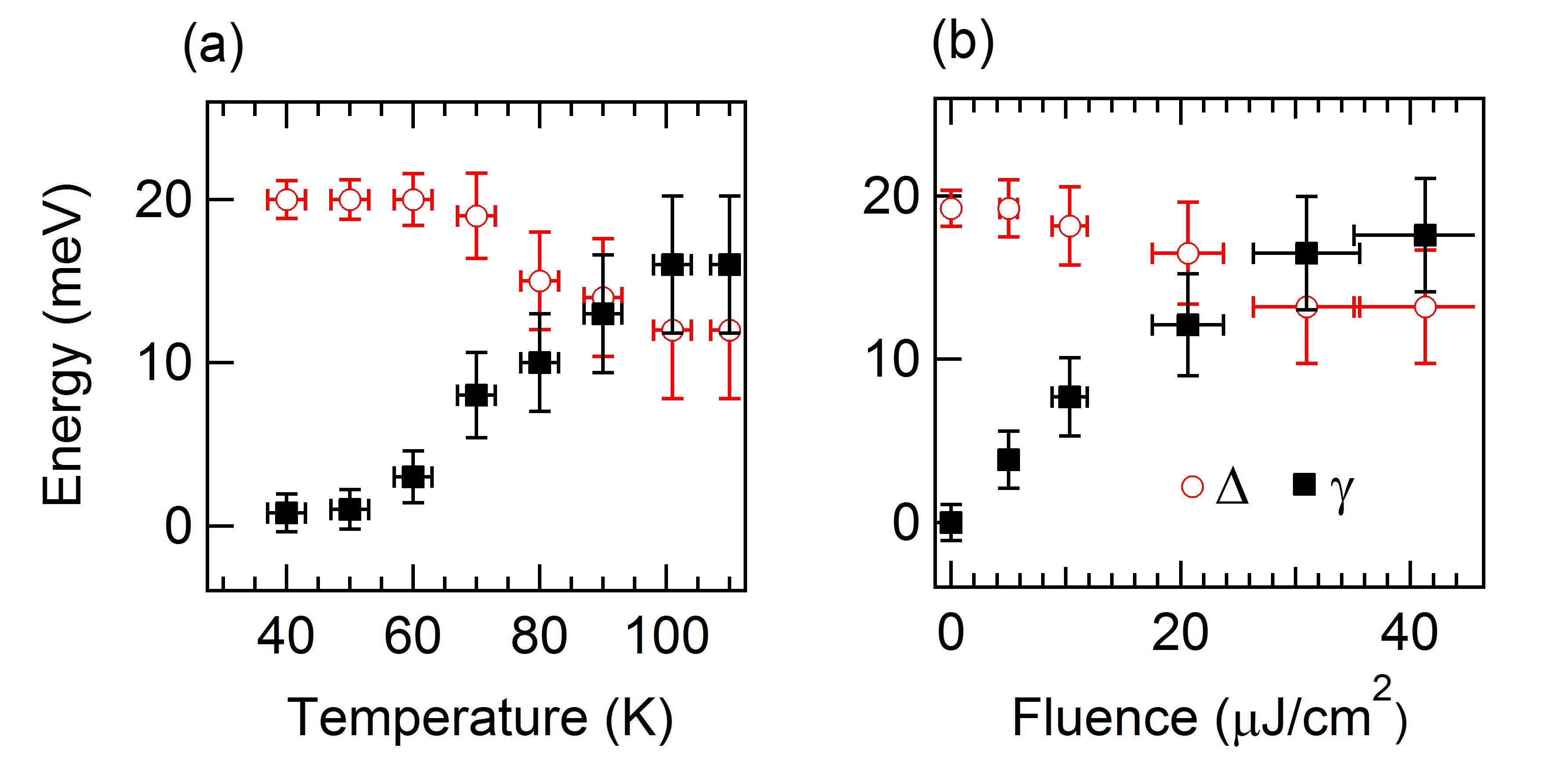} 
\caption{Superconducting gap $\Delta$ (red circles) and the pair-breaking rate $\gamma$ (dark squares) as a function of temperature (a) and photoexcitation fluence (b). These parameters have been extracted by fitting the symmetrized EDCs in Fig. \ref{Fig2}(f) and Fig. \ref{Fig3}(d).}
\label{Fig4}
\end{center} 
\end{figure}

\begin{figure}
\begin{center}
\includegraphics[width=0.95\columnwidth]{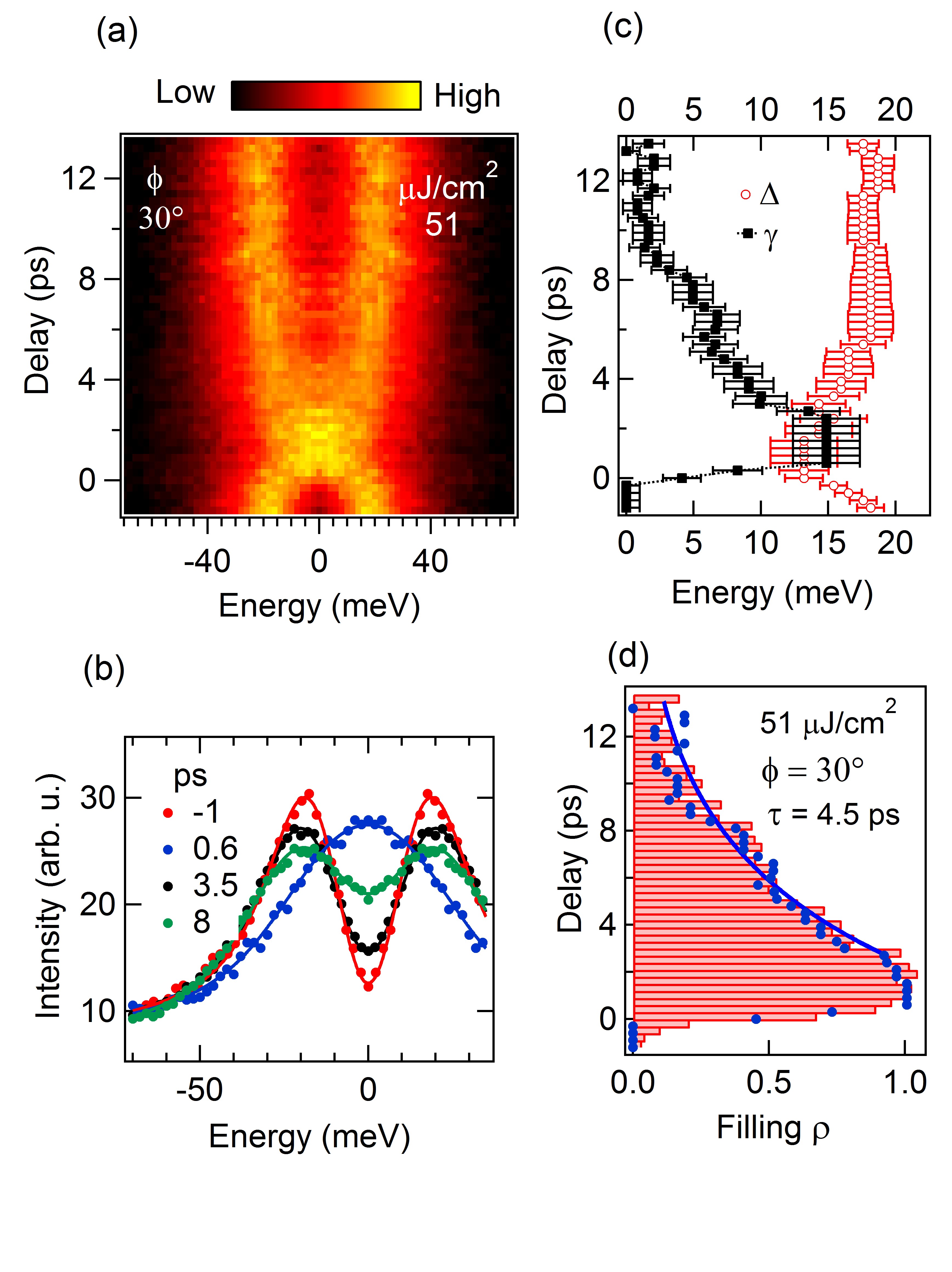}
\caption{a) Intensity map of symmetrized EDCs as a function of pump probe delay. The data have been acquired at 40 K, $\phi=30^\circ$ and pump fluence $F = 51 \mu J/cm^2$. b) Symmetrized EDCs (marks) extracted at selected pump probe delays are compared with the fitting curves (solid lines). c) Temporal evolution of the gap $\Delta$ (red circles) and of the pair-breaking rate $\gamma$ (dark squares). (d) the temporal evolution of the gap filling factor $\rho$ (filled bars) is compared with $1.8/(1+\Delta/\gamma)$ (blue circles). The solid line is an exponential fitting function with time constant $\tau=4.5$ ps.}
\label{Fig5}
\end{center}
\end{figure}

The description of the non-equilibrium case is first performed by acquiring photoelectron intensity maps at intermediate pumping fluences.  We observed that photoexictation also generates a photoinduced band displacement smaller than 3 meV. The quantitative estimate of such rigid shift is obtained by fitting momentum distribution curves at energy below -70 meV. Although the physical origin of the spectral shift is still debated \cite{Smallwood_Shift, Bovensiepen_Shift}, we suspect the occurrence of photoinduced fields at the sample surface. In the following, this minor correction has been considered when constructing the symmetrized EDCs. Figure \ref{Fig3}(c) shows EDCs extracted at the Fermi wavevector for increasing pumping fluence. Also in this case, the symmetrized EDCs of Figure \ref{Fig3}(d) are accurately fitted by the model spectral function. 

We compare in Fig. \ref{Fig4} the fitting parameters of the curves in Fig. \ref{Fig2}(f) (thermal heating) and Fig. \ref{Fig3}(d) (sudden photoexcitation). In both case the collapse of the superconducting phase is dominated by the increase of the pair-breaking rate $\gamma$. Once $\gamma>\Delta$ the binding energy of Cooper pairs is ill defined and the incertitude of $\Delta$ becomes increasingly large. Note in Fig. \ref{Fig4}(a) that $\gamma$ is appreciably different from zero only if the temperature is larger than $0.7T_c\cong 60$ K. We evince that superconducting fluctuations are thermally activated near to the transition temperature. On the other hand, even photon pulses with arbitrarily small fluence can generate a detectable pair-breaking. In the weak perturbation regime, the density of the photoinduced fluctuations is proportional to the number of absorbed photons. As a consequence the pair-breaking rate in Fig. \ref{Fig4}(b) displays a linear scaling for $F<20 \mu$J/cm$^2$.

The dynamics of gap filling is captured by acquiring photoelectron intensity maps at variable pump probe delays. Figure \ref{Fig5}(a) shows on a false color scale the temporal evolution of the symmetrized EDCs while Fig. \ref{Fig5}(b) display symmetrized EDCs and fitting curves for selected delays. At 0.6 ps, the photoexcitation by $51 \mu J/cm^2$ results in the complete filling of the superconducting gap. We show in Fig. \ref{Fig5}(c) the temporal evolution of the parameters $\Delta$ and $\gamma$ as a function of delay time. The near nodal gap is fully melted as long as $\gamma>\Delta$, namely during the first 2 picoseconds. For practical purpose, it is useful to quantify the melting of superconducting phase by the filling factor $\rho$. This phenomenological observable is independent on the specific model and saturates in the regime of the collapsed gap. We obtain the filling factor by integrating the symmetrized EDCs in an energy window of 20 meV around the Fermi level and linearly rescaling the resulting values so that $\rho$ is comprised in the interval $[0,1]$. Figure \ref{Fig5}(d) shows that the evolution of $\rho$ nearly follows $1.8/(1+\Delta/\gamma)$. Therefore, the increase of $\rho$ depends both on the gap shrinking and on $\gamma$. The filling factor is correctly fit by an exponential function with recovery time of $4.5 \pm 0.5$ ps. Our data are in qualitative in agreement with the work of Smallwood \emph{et al.} \cite{Smallwood_gap}. However, the analysis of ref.  \cite{Smallwood_gap} does not consider the finite pair-breaking rate. Therefore, the fitting parameters cannot be directly compared. We outline here that superconducting models with finite $\gamma$ have a natural link to established interpretations of the transient optical measurements\cite{Kabanov,Demsar_THz}. Indeed, it has been often proposed that Cooper pairs are broken by the inelastic scattering with the non-equilibrium phonons that follow from the photoexcitation process. Within this framework, the recovery of superconductivity takes place via the heat transfer towards less harmful modes and has been successfully simulated by Rothwarf-Taylor equations \cite{Kabanov}.

\section{Spectroscopy of the gap away from the nodal direction}\label{sec5}

\begin{figure}
\begin{center}
\includegraphics[width=1\columnwidth]{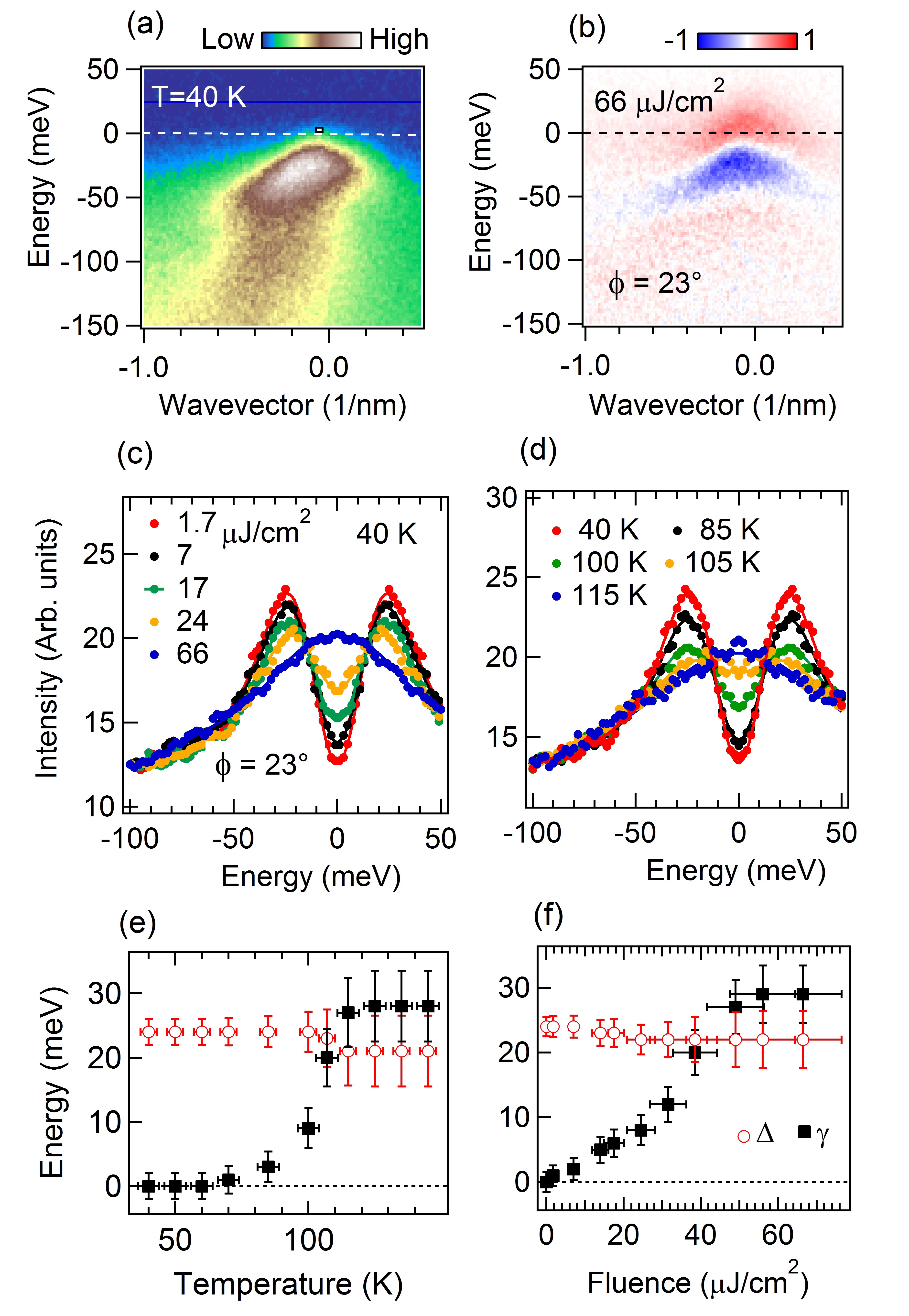}
\caption{The data of this image have been acquired at azimutal angle $\phi=23^\circ$. a) Intensity map in equilibrium at the base temperature of 40 K. b) Pump-on minus pump-off signal acquired at 40 K with photoexcitation fluence $F= 66 \mu J/cm^2$. c) Symmetrized EDCs acquired at 40 K and just after photoexcitation with different pumping fluence. Fitting curves are superimposed to the experimental data d) Symmetrized EDCs (marks) acquired in equilibrium conditions at different sample temperatures. Fitting curves (solid lines) are superimposed to the experimental data. Evolution of gap $\Delta$ (red circles) and pair-breaking rate $\gamma$ (dark squares) with temperature (e) and photoexcitation fluence (f).}
\label{Fig6}
\end{center}
\end{figure}

\begin{figure}
\begin{center}
\includegraphics[width=1\columnwidth]{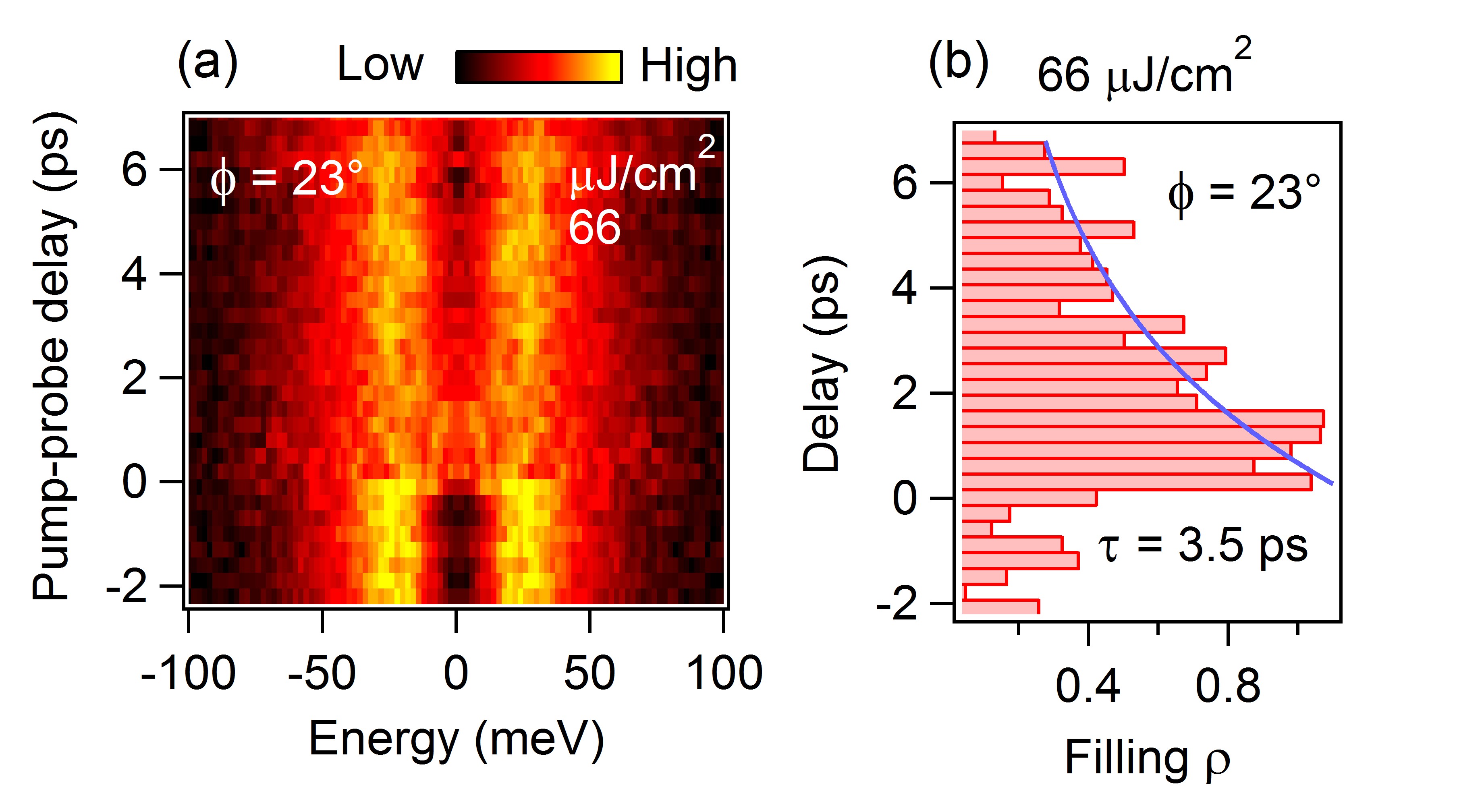}
\caption{a) Intensity map of symmetrized EDCs as a function of pump probe delay. The data have been acquired at 40 K, $\phi =23^\circ$ and pump fluence $F= 66 \mu J/cm^2$. b) Temporal evolution of the gap filling factor and exponential function with time constant $\tau=3.5$ ps. The filling factor is extracted from the simmetrized EDCs by integrating the spectral intensity in an in an energy window of 20 meV centered at the Fermi level.}
\label{Fig7}
\end{center}
\end{figure}

Figure \ref{Fig6}(a) displays a photoelectron intensity map acquired at azimuthal angle $\phi=23^\circ$ and 40 K. Due to bilayer splitting, the map contains two parallel QP bands that are not spectrally resolved. A strong renormalization of QP dispersion takes place for excitation energy above -70 meV. The presence of the superconducting gap induces the weak backfolding of dispersive peaks at the Fermi wavevector. We show in Fig. \ref{Fig6}(b) the pump-on minus pump-off map acquired just after optical pumping with $66 \mu J/cm^2$. The observed contrast visualizes the transfer of spectral weight from the QP peak to the gapped spectral region. Symmetrized EDCs of Fig. \ref{Fig6}(c) display the photoinduced filling of the gap at different pumping fluence. Any signature of pairing is lost at high photoexcitation density. We show in Fig. \ref{Fig6}(d) that a similar spectral evolution takes place in equilibrium by increasing the sample temperature. Simulations based on the intermediate coupling model have been superimposed to the experimental EDCs. The partially filled gap above $T_c$ is a signature of superconducting fluctuations \cite{Alloul, Bergeal, Ong, Perfetti_THz}. At $\phi=23^\circ$ the condition $\Gamma>\Delta$ takes place at $T\cong 1.3 T_c$ so that preformed Cooper pairs exist for $T_c<T<1.3T_c$. Note that such gradual filling of the gap along the off nodal direction cannot be captured by models neglecting the pair-breaking \cite{Smallwood_gap}.  Figure \ref{Fig6}(e) shows that $\gamma$ increases appreciably only above 60 K. This results is consistent with the superconducting phase transition being ruled by thermal fluctuations in a restricted temperature window around $T_c$.
However, the sudden photoexcitation generates superconducting fluctuations via a distinct mechanism. Figure \ref{Fig6}(f) confirms that $\gamma$ scales linearly at low pumping fluence. This finding is the signature of non-thermal processes leading to pair-breaking. Beside the purely electronic excitations, fluctuations can also originate from the high average occupation of harmful phonons. The emergent role of such hot phonons would also explain the large discrepancy between the threshold energy for a photoinduced phase transition and the condensation energy measured in thermal equilibrium \cite{Demsar_THz}. Finally, figure \ref{Fig7}(a) shows the temporal evolution of symmetrized EDCs at $\phi=23^\circ$. The dynamics of filling factor in Fig. \ref{Fig7}(b) indicates that the recovery of the gapped phase takes place on the characteristic timescale of $3.5 \pm 0.7$ ps. Within the error bars, this value is consistent with the one obtained at $\phi=30^\circ$.
 
\section{Gap evolution and quasiparticles dissipation}\label{sec6}

\begin{figure}
\begin{center}
\includegraphics[width=1\columnwidth]{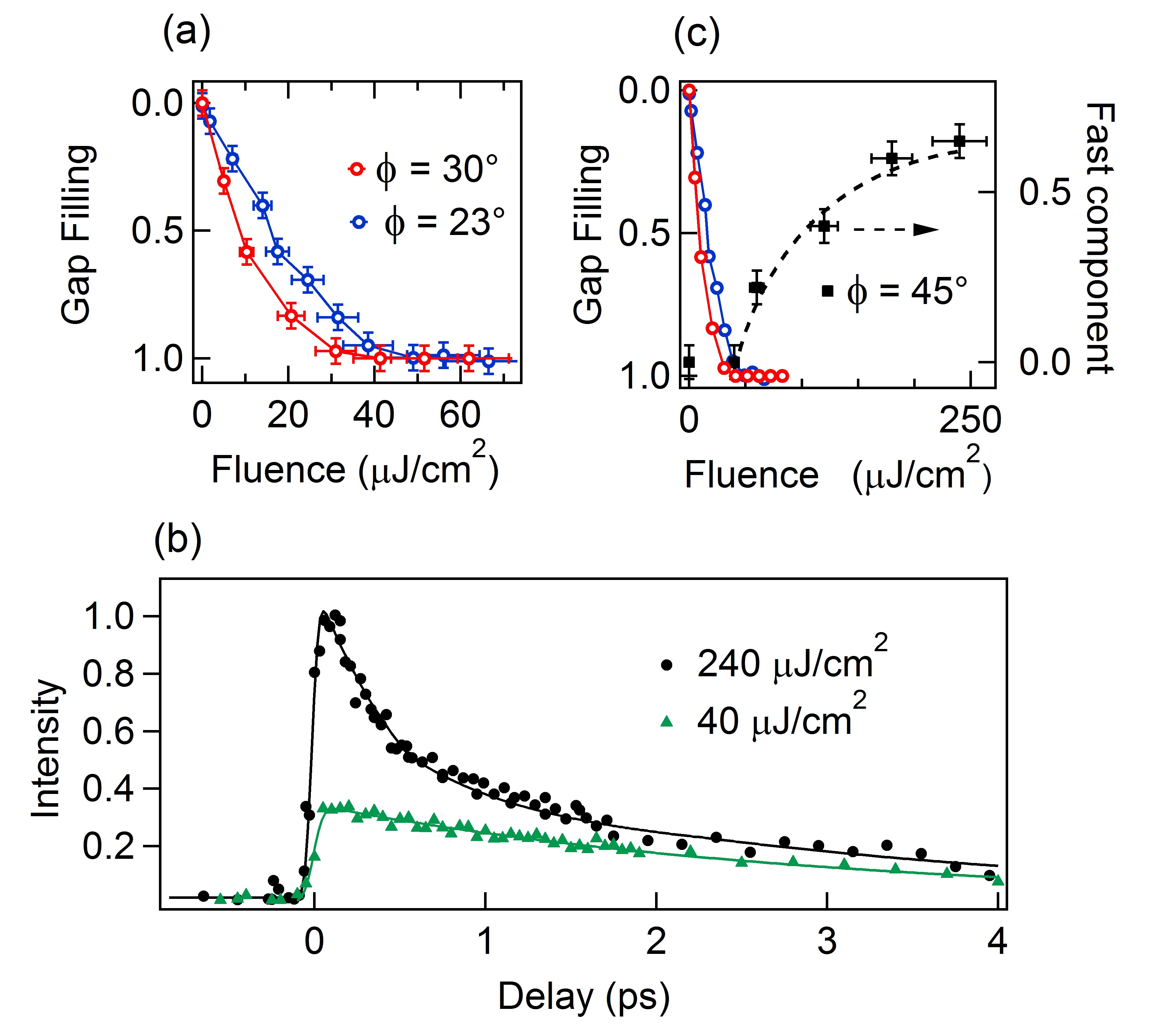}
\caption{a) Filling factor of superconducting gap at $\phi=30^\circ$ (red circles) and $\phi=23^\circ$ (blue circles) extracted at roughly 300 fs after photoexcitation with a variable pump fluence. b) Temporal evolution of nodal quasiparticle intensity integrated in an energy window from the Fermi level up to excitation energy of 50 meV. Comparison between the QP relaxation after photoexcitation with $40 \mu J/cm^2$ (green triangles) and $240 \mu J/cm^2$ (black circles). c) The filling factor of the gap (open circles) and the fast component of the nodal QP relaxation (dark squares).}
\label{Fig8}
\end{center}
\end{figure}

We show in  Fig. \ref{Fig8}(a) the filling factor of the superconducting gap as a function of pump fluence acquired at azimuthal angles  $\phi=30^\circ$ and $\phi=23^\circ$. Our data extend the work of Smallwood \emph{et al.} \cite{Smallwood_gap} and show that no signature of pairing can be detected above $40 \mu J/cm^2$. By this means we exclude that spectra at $\phi\geq 23^\circ$ develop a pseudogap surviving up to high photoexcitation density. Note in Figure \ref{Fig8}(a) that the filling factor of the superconducting gap depends slightly on the azimuthal angle. For a given fluence, the relative weight of in-gap states is larger towards the node ($\phi=30^\circ$) than farther for it ($\phi=23^\circ$). The anisotropic character of gap melting is common both to the photoinduced and to the thermally heated state \cite{Shin}. This finding is directly linked to the $d$-wave symmetry of the superconducting gap, leading to $\gamma>\Delta$ at larger fluence (or temperature), when moving off the nodal direction.

Next we compare the temporal evolution of the gap with the energy relaxation of nodal QP. In the latter case we employ the setup providing 80 fs probe pulses with 70 meV bandwidth.
Figure \ref{Fig8}(b) shows the temporal evolution of nodal QP signal integrated in the energy window [0,50] meV. After photoexcitation with fluence $F= 40 \mu J/cm^2$ the QPs follow a single exponential relaxation with decay constant of 2.5 ps. The relaxation of hot electrons becomes qualitatively different when pumping the sample with more intense pulses. Upon photoexcitation with $F=240 \mu J/cm^2$ the dynamics displays an initial decay with inverse rate of 0.5 ps. At longer delays the cooling time converges to 2.5 ps, independently on photoexcitation intensity. These results are consistent with the data first reported by Cortes \emph{et al.} \cite{Bovensiepen_QP} and have been thoroughly discussed in ref. \cite{Piovera}. Phenomenologically, the relaxation of the energy integrated QPs signal can be correctly reproduced by a double exponential decay. The fast component becomes visible above a threshold fluence and gains weight upon increasing the photoexcitation density. We compare in Fig. \ref{Fig8}(c) the amplitude of the fast QPs delay with the filling factor of the superconducting gap. The fast dissipation rate becomes detectable at the same threshold fluence where the superconducting gap has fully collapsed. Such connection has been already outlined by ref. \cite{Smallwood_QP} although at fluence values lower than the ones of our work. Smallwood \emph{et al.} proposed that a dynamical gap opening affects the recovery rate of photoexicted quasiparticles \cite{Smallwood_Model}. In this case the reforming pairs not only hinder quasiparticle dissipation but also act as an heat bath for nodal QPs. The energy released by Cooper pairs formation can be transferred effectively to QPs, drastically reducing their cooling time. We support this plausible scenario, which is likely valid also in the intermediate coupling regime.

\section{Conclusions and Acknowledgments} In conclusion our time resolved ARPES data of Bi$_2$Sr$_2$CaCu$_2$O$_{8+\delta}$ compare the collapse of the superconducting gap upon adiabatic heating and sudden photoexcitation. The intermediate coupling model correctly accounts for the gap filling, both near and off the nodal direction. Within this framework, the melting of the superconducting phase takes place when the pair-breaking rate overcomes the electronic gap. Interestingly, the superconducting fluctuations are thermally activated in equilibrium conditions whereas they scales linearly in the weak photoexcitation regime. This finding suggest that non equilibrium phonons are the main source of pair-breaking in the transient state. The data in the off nodal direction report a full collapse of the gap for $F>40 \mu$ J/cm$^2$ and exclude the presence of a  pseudogap unrelated to superconductivity for $\phi>23^\circ$. Finally, we compare the narrow bandwidth spectroscopy of the gap, with a short pulses spectroscopy of the QP relaxation. The latter develops a faster component at the threshold fluence where the gap has fully collapsed. Dynamical reformation of the superconducting gap is a plausible explanation for such dynamics.

This work is supported by "Investissements d'Avenir" LabEx PALM (grant No. ANR-10-LABX-0039PALM), by the China Scholarship Council (CSC, Grant No. 201308070040) by the EU/FP7under the contract Go Fast (Grant No. 280555), and by the R\'egion Ile-deFrance through the program DIM OxyMORE. We are thankfull to Andrea Gauzzi for critical reading and discussing the manuscript.

\end{document}